%% file: BRZ-PRL-v2.tex
\documentclass[twocolumn,nofootinbib]{revtex4-1}
\usepackage{latexsym,epsfig,amssymb, amsmath,nicefrac}

\usepackage[makeroom]{cancel}
  \usepackage{latexsym}
  \usepackage{epsf}
  \usepackage{amssymb}
  \usepackage{graphicx}
  \usepackage{amsmath}
  \usepackage{amsmath,amssymb,amsthm}
  \usepackage{verbatim}
  \usepackage{hyperref}

\newcommand{\w}{\wedge}

\def\bi{\begin{itemize}}
\def\ei{\end{itemize}}
\def\be{\begin{equation}}   
\def\ee{\end{equation}}
\newcommand{\bea}{\begin{eqnarray}}
\newcommand{\eea}{\end{eqnarray}}

\begin{document}

\title{{\Large De Sitter Vacua from Non-perturbative Flux Compactifications}}

\author{ Johan Bl{\aa}b\"ack$^1$, Diederik Roest$^2$, Ivonne Zavala$^2$}

\affiliation{{}$^1$Institutionen f{\"o}r fysik och astronomi, Uppsala Universitet,\\ Box 516, SE-751 08 Uppsala, Sweden, johan.blaback@physics.uu.se}

\affiliation{{}$^2$Centre for Theoretical Physics, University of Groningen, \\ Nijenborgh 4, 9747 AG Groningen, The Netherlands, $\{$d.roest, e.i.zavala$\}$ @ rug.nl}

\begin{abstract}
We present stable de Sitter solutions of $\mathcal{N} = 1$ supergravity in a geometric type IIB duality frame with the addition of non-perturbative contributions. Contrary to the standard approach, we retain the moduli dependence of both the tree level superpotential and its non-perturbative contribution. 
 This provides the possibility for a single-step  stabilisation of all moduli simultaneously in a de Sitter vacuum.  Using a genetic algorithm we find explicit  solutions with different features.

\end{abstract}

\maketitle

\smallskip

\noindent  {\bf Introduction.}
The  importance of accelerating space-times in cosmology, both for inflation and dark energy, makes it critical to understand the role of de Sitter (dS)  vacua in string theory. Many such constructions  have been criticised as being rather ad-hoc. In the KKLT set-up \cite{KKLT}, one adds a non-perturbative contribution as well as explicit, supersymmetry-breaking uplift terms to achieve a dS vacuum. These are necessary additions to $\mathcal{N} =1$ compactifications with only IIB gauge fluxes, which only lead to Minkowski vacua with flat directions \cite{GKP}. On the IIA side, the situation regarding moduli stabilisation is  better, as the inclusion of  gauge fluxes alone leads to AdS vacua \cite{DeWolfe}. However, it is not possible to obtain dS  solutions in this vein \cite{Hertzberg}. Adding metric-fluxes does lead to dS solutions \cite{Munich,Flauger:2008ad}, but all known examples are unstable. 
In this Letter,  we show that geometric and isotropic fluxes with non-perturbative contributions are enough to stabilise simultaneously  all moduli in a dS vacuum, in the simplest scenario possible, widening the dS landscape. 

\smallskip
We focus on a $T^6/(\mathbb{Z}_2 \times \mathbb{Z}_2)$ compactification with  fluxes in type IIB supergravity in ten dimensions. 
The number of untwisted moduli is  $(h^{(1,1)},h^{(2,1)}) =(3,3)$ plus the dilaton. We concentrate on the isotropic case with a single K\"ahler and complex structure moduli, that is, an $STU$-type of model. The K{\"a}hler potential takes the form: 
\begin{equation}
K = -\log[-i(S-\bar{S})] - 3 \log[-i( T-\bar{ T})] - 3 \log[-i(U-\bar{U})].
\end{equation}

\noindent The scalar potential takes the usual form (we are setting $M_P^{-2}=8\pi G =1$):
\begin{equation}
V = e^{K} \left( K^{I\bar J}D_IWD_{\bar J}\overline{W} - 3 |W|^2 \right),
\end{equation}
$D_IW= \partial_I W + \partial_I K W $ with $I, J$ labelling all moduli.

The tree-level superpotential depends on the dilaton $S$ and complex structure $U$ (jointly 
referred to as complex structure moduli). These are generated  by  the presence of RR-flux $F_3$ and  NSNS-flux $H_3$ (with coefficients $a_i$, $b_i$, respectively):
 \be \label{W0}
  W_{tree} = P(a_i,U) - S P(b_i,U)\,,
 \ee
where $P(f_i,U)$ are polynomials in $U$ of the form
 \begin{equation}\label{Ps}
  P(f_i, U) = f_0 - 3 f_1 U + 3 f_2 U^2 - f_3 U^3 \,.
 \end{equation}
Thus, these fluxes generate a potential for the complex structure moduli stabilising them \cite{GKP}. However, the K\"ahler modulus remains as a flat direction.

To stabilise the $T$-modulus, the standard approach has been to first use the tree-level flux contributions to fix $S$ and $U$  in a SUSY vacuum. Second,  introduce a non-perturbative term  $W_{\textrm{NP}}(T)$ for the K\"ahler modulus, allowing its stabilisation.   It is assumed that the first step results in a constant contribution to the superpotential, $W_0=$const. and a constant coefficient for the non-perturbative term, $A_0=$const.:
\be
W = W_0 + A_0 \, e^{i x T}\,,
\ee
where $x = 2\pi/ K$ for gaugino condensation with gauge group rank $K$. Using this superpotential, only AdS minima can be obtained. Therefore,  a final third step has been taken by adding a suitable {\em uplifting} term \cite{KKLT}, lifting the AdS minimum to a dS. 

This three-step process  has been criticised in the literature since in general, not only $W_0$  but also  the coefficient $A_0$ depend on the complex structure moduli \cite{Witten:1996bn}.  Therefore the second step can lead to complications since heavy modes could mix with light modes \cite{deAlwis:2005tf,Shanta1} and create instabilities \cite{Choi:2004sx} (see however \cite{Gallego} for a discussion on the consistency conditions for this step). 
Moreover, adding an uplifting term by hand is  under limited theoretical control (e.g.~adding an anti-brane is an explicit heavy breaking of  supersymmetry and it is  not clear that one can still use a supergravity description). An alternative was proposed to uplift using  D-terms in  \cite{Burgess:2003ic, Achucarro:2006zf}.
Finally, the third step has been relaxed in \cite{Marta1}, where it was   shown how to obtain 
dS minima without the need of an artificial uplifting term\footnote{Further uplifting alternatives using perturbative corrections  have been discussed in \cite{Fer,Louis}.}.
However the second step has been so far assumed  in order to obtain stabilisation of all 
 moduli\footnote{A one-step stabilisation in  heterotic orbifold compactifications using the explicit modular covariance of the superpotential has been discussed in \cite{Parameswaran:2010ec}. In  the  Large Volume scenario  \cite{Balasubramanian},  an alternative single-step stabilisation giving  rise to dS vacua,  has been achieved term by term in an expansion in inverse powers of the volume \cite{LV1,LV2,LV3}. These solutions include perturbative corrections to the K\"ahler potential, 
  D-terms and chiral matter,   and thus go beyond our present discussion. }. This Letter addresses the natural question whether a single-step process can give rise to (meta-)stable dS vacua,  stabilising all moduli simultaneously, even in simple models\footnote{A similar approach in the IIA duality frame is discussed in \cite{Danielsson:2013XX}.}. \\

\noindent {\bf A novel mechanism of moduli stabilisation.}
To study this  possibility, we consider the usual tree-level superpotential augmented with a non-perturbative term. The coefficient of the latter will generically depend on the complex structure moduli $S$ and $U$ in a non-trivial but unknown way. Following the reasoning for non-geometric fluxes of \cite{Wecht}, it seems natural to model this dependence in a way that respects the $S$ and $U$ duality covariance\footnote{Indeed in a duality covariant formulation of ${\cal N} = 1$ supergravity with non-geometric fluxes, it is possible to find  stable dS solutions \cite{Guarino}. }.
This leads to the following Ansatz:
\begin{equation}\label{W1}
W = W_{tree} + \left[P(\tilde{a}_i,U) - P(\tilde b_i, U)S\right]  e^{i x T},
\end{equation}
where all four polynomials have the structure \eqref{Ps}. Although 
we do not provide a rigorous derivation of this moduli  dependence in the non-perturbative term, it is constrained to be such a polynomial in $S, U$  by duality arguments; indeed a similar form has been studied in an explicit string theory setting \cite{Uranga}. An alternative interpretation of this Ansatz is that the polynomials represent a Taylor expansion  in terms of small $S, U$, as we will justify below.

The coefficients of the non-perturbative term appear in complete analogy with the gauge fluxes in the superpotential. Thus   
 we  refer to $\tilde{a}_i$ and $\tilde{b}_i$ as non-perturbative fluxes. This leads to a total set of 16 fluxes. However, as we explain later,  it will be sufficient to have 12 fluxes. We  therefore set the fourth polynomial  equal to zero.

\smallskip
To find solutions, we  use the property that any solution to the equations of motion can be represented by a solution in the origin of moduli space (in our conventions located at $S=T=U=i$ with $x=1$). This technique was first proposed in the context of half-maximal supergravity \cite{Dibitetto:2011gm} and subsequently used to explore the vacuum structure of maximal supergravity \cite{DallAgata, Borghese}, but can be applied to any theory with a homogeneous scalar manifold. This avoids an over-counting of solutions and reduces dramatically  the complexity of the equations of motion. While these in general can be high-degree polynomials for the fields, in the origin these reduce to quadratic equations in terms of fluxes. Solutions correspond to flux configurations for which these quadratic combinations vanish. 
The origin is however not a configuration that should be considered a valid supergravity limit, since the volume of the internal space and the string coupling are both equal to one. Below we explain how to deal with this issue.

To solve  the resulting  quadratic equations in the fluxes $\{a_i,b_i,\tilde{a}_i\}$, we  use the fact that when supersymmetry is preserved, the  equations of motion are implied \cite{Danielsson:2012by}:
\begin{equation}\label{eq:SBP}
D_I W \equiv A_I + i B_I = 0 \quad \Rightarrow \quad \partial_I V = 0,
\end{equation}
where the six supersymmetry breaking (\cancel{SUSY}) parameters $A_I$ and $B_I$ are linear combinations of the superpotential couplings $\{a_i, b_i,\tilde a_i \}$. It will be advantageous to split up the latter in (linear combinations of) two sets: there are six \cancel{SUSY} parameters while the orthogonal combinations preserve SUSY. Via this approach, 
in general all moduli take part in \cancel{SUSY} and contribute to the uplifting of the potential. 
This is to be contrasted to for example \cite{KKLT}, where uplifting is only considered in the direction of Im($T$) while $S$ and $U$ are in SUSY minima.

Next, one exploits the fact that the equations of motion are implied by SUSY. Therefore, the equations of motion become quadratic in the \cancel{SUSY} parameters or bilinear in the \cancel{SUSY} and SUSY parameters. 
For this to work, the total set of parameters must  be at least equal to twice the number of (real) fields, in our case $6+6=12$. Type IIB tree-level flux contributions to the superpotential consists of $8$ parameters $a_i,b_i$  (\ref{W0}, \ref{Ps}). In \cite{Danielsson:2012by} the extra couplings were taken to be so called non-geometric fluxes. Here, we add the non-perturbative fluxes $\tilde a_i$ (\ref{W1}).

Given these sets of solutions parameterised by the six \cancel{SUSY} parameters, we follow \cite{Blaback:2013ht} in using a genetic algorithm to scan this parameter space to look for stable dS solutions
(for similar applications of genetic algorithms see \cite{Damian:2013dq,Damian:2013dwa,Blaback:2013fca}). 
 We thus require both the cosmological constant as well as all the scalar masses, obtained by diagonalising the mass matrix
\begin{equation}\label{massmatrix}
(m^2)^I_J = \frac{K^{IK} \partial_K \partial_J V}{V} \,,
\end{equation}
to be positive. \\

\noindent {\bf Perturbative reliability}.
In order to get to a regime for the values of the moduli that is a reliable supergravity approximation of string theory, we 
need to ensure that we work at large volume and small string coupling, such that higher string mode contributions and loop corrections are suppressed:  a) Large volume: ${\cal V}\sim  r^6 \gg 1$\,.
b) Small string coupling: $g_s^{-1}  \gg 1$,  where ${\cal V}= \textrm{Vol.}/ \ell_s$ with $\ell_s$ the string scale, and we have introduced a characteristic (dimensionless) radius of the internal space, $r$. The volume is further given in terms of the overall K\"ahler modulus as 
${\rm Im}\, T = {\cal V}^{2/3}$ and the string coupling in terms of the dilation as  $g_s=({\rm Im} \,S)^{-1}$.

Consider the following rescaling of the volume and the string coupling $r \to N^\alpha r ,\quad g_s \to N^{-\beta} g_s,$ for $\alpha$ and $\beta$ some positive numbers with  $N \gg 1$. From the expression for the scalar potential, the fluxes and $x$ have to be rescaled as:
\begin{align}
 a_i,\tilde{a}_i & \to N^{6 \alpha + \beta/2 + \gamma} \, a_i,\tilde{a}_i \,,   \qquad x\to N^{-4 \alpha }\, x \,, \notag \\
 b_i & \to  N^{6\alpha -\beta/2 + \gamma}\, b_i \,, \label{xscale1}
\end{align}
for the  solution to be  preserved. We have also introduced a parameter $\gamma$ that represents an overall scaling of the fluxes that is always possible to perform. The potential scales as $V \to N^{2\gamma} \, V \,,$ and the normalised masses remain invariant.
For the special case of gaugino condensation, where $x=\frac{2\pi}{K}$, the scaling (\ref{xscale1}) implies that we need to scale $K$ as:  $K\to N^{4 \alpha }\, K\,.$

Given a solution, we can  achieve a large volume and small string coupling regime, via a suitable rescaling of the parameters. A drawback may be that this rescaling requires a small value of the parameter $x$, which  in the case of gaugino condensation, translates into a large rank of the gauge group $K$. 
 In the context of non-compact Calabi-Yaus, it has been discussed that arbitrarily high gauge group ranks are possible \cite{GVW}.  
 In the compact case the situation turns out to be more restrictive, but relatively large values are possible \cite{Louis}. 

Finally, we should also consider the tadpole cancellation condition, which is a quadratic combination of flux parameters, $H_3 \w F_3$:
 \begin{equation}
  N_{\textrm{D}3} = a_3 b_0 - 3 a_2 b_1 + 3 a_1 b_2 - a_0 b_3,
 \end{equation}
scaling according to   $N_{\textrm{D}3} \to N^{12\alpha + 2\gamma} \, N_{\textrm{D}3}.$
As the tadpole is bounded from below by the orientifold contribution, one should worry about this rescaling in the case of negative $N_{\textrm{D}3}$. Indeed, in all our examples below, the tadpole will be negative. In order to avoid that the large volume limit pushes the tadpole below its lower limit, one can  choose the $\gamma$ parameter suitably.

Notice that there is no particular requirement of the value for the complex structure modulus $U$ at the minimum. Therefore,  we keep this field to the origin. However, we could rescale it as well to small values in such a way that the power expansion in the non-perturbative function $P_3$ can be truncated at the third power consistently.

We next consider the relevance of possible perturbative corrections to the  K{\"a}hler potential since these could dominate over the non-perturbative contributions to the superpotential, rendering the present set-up inconsistent. Perturbative contributions scale with $K_{\rm P} \sim 1/({\cal V}g_s^{3/2})$.
Since our method to find solutions starts with all fields at the origin and fluxes of the same order, all contributions in the superpotential are of the same order. After making  the above described rescalings, all terms in the potential scale in the same way and once perturbative K\"ahler contributions are added, we can write 
\begin{equation}
V_\textrm{full} \to  N^{2\gamma} \left( V + K_\textrm{P} V \right) \sim N^{2\gamma} \left( V + N^{-6 \alpha+\frac{3\beta}{2}} V \right),
\end{equation}
hence $K_\textrm{P}$ contributions will be suppressed with a factor $N^{- 6 \alpha+\frac{3\beta}{2}}$ compared to the potential calculated here. We can therefore safely neglect these  by a suitable choice of $\alpha, \beta$. \\

\noindent {\bf Explicit de Sitter solutions.}
We performed five individual searches where additional criteria were required.
These additional criteria were chosen to be:

$1,2:$ maximize and minimize 
$\tilde{\gamma}=|DW|^2/(3|W|^2)$ (while keeping $\tilde{\gamma} > 1$),

$3,4:$ maximize and minimize the scale between the fluxes $|b_i|/|a_i|$,

$5:$ minimize the scale between the fluxes $|\tilde{a}_i|/|a_i|$.\\

\begin{table}[htbp!]\tiny
\centering
\input{./new_sol_table_Bert.tex}
\caption{Properties of the solutions. The masses are normalised with the potential and all 
scales are given in Planck units. 
\label{tab:sols}}
\end{table}

The main properties of our solutions are presented in Table \ref{tab:sols}, while the \cancel{SUSY} parameters for these solutions can be found in Table \ref{tab:SBP}. A number of general features can be extracted from these examples.
Firstly, it follows that the \cancel{SUSY} and AdS scales are always of the same order and cannot be separated. The maximum ratio between the scales is $\approx 1.0256$, as follows from solution 1. Similarly, one can approach a ratio equal to one with very good accuracy, as illustrated by solution 2. Another observation  from this solution is that the lowest mass can be made very large compared to the potential energy $V_0$. Finally, as the flux parameters are of decreasing order,
\begin{equation}
\{ \tilde{a}_0, \tilde{a}_1, \tilde{a}_2, \tilde{a}_3\} \approx \{ 0.0835,0.0702 ,0.0372 ,0.00921 \}
\end{equation}
the small-$U$ expansion of \eqref{W1} is justified in this case.

A second point is that we tried to achieve a hierarchy between the RR- and NSNS-fluxes. The reason for doing so is the small coupling limit; as the rescaling \eqref{xscale1} acts different on these two set of fluxes, we would like to start off with a hierarchy of values for these. After the rescaling we  end up at small coupling with fluxes of the same order. As can be seen from solutions 3 and 4, it is possible to achieve a small degree of separation between the two sets of fluxes. However this separation is only due to large contributions from the non-perturbative fluxes. 

Finally, in solution 5 we were able to create a hierarchy between the perturbative and  non-perturbative fluxes. This hierarchy is  only possible to achieve with the loss of a hierarchy of the NSNS and RR sector. The reason for this lies in the structure of the equation of motion for $S$. On the level of the perturbative superpotential this equation forces the so-called imaginary self-dual (ISD) condition for the flux $G_3 = F_3 + S H_3$ \cite{GKP}. Via the addition of small non-perturbative contributions it is only possible to perturb this condition. This is why we see in solutions 3 and 4 that the non-perturbative terms contribute much more than in solution 5. For the same reason, i.e.~small non-perturbative contributions cannot significantly change the ISD condition,  we are not able to find solutions without net O-planes, as is argued to be possible in type IIA \cite{Danielsson:2013XX}.

For the most interesting solution 5, we observe also that because the non-perturbative contributions are suppressed, 
 one may expect a separation of masses among $S,U$ and $T$. Indeed the last two smallest masses in Table \ref{tab:sols} correspond to the eigenvectors which are dominated by the real and imaginary parts of $T$. The other small mass corresponds mostly to a combination of the $S,U$ axions. 
Moreover, the lowest mass is still significantly larger than the potential.  

\begin{figure}[htbp!]
\centering
\includegraphics[scale=.38,keepaspectratio=true]{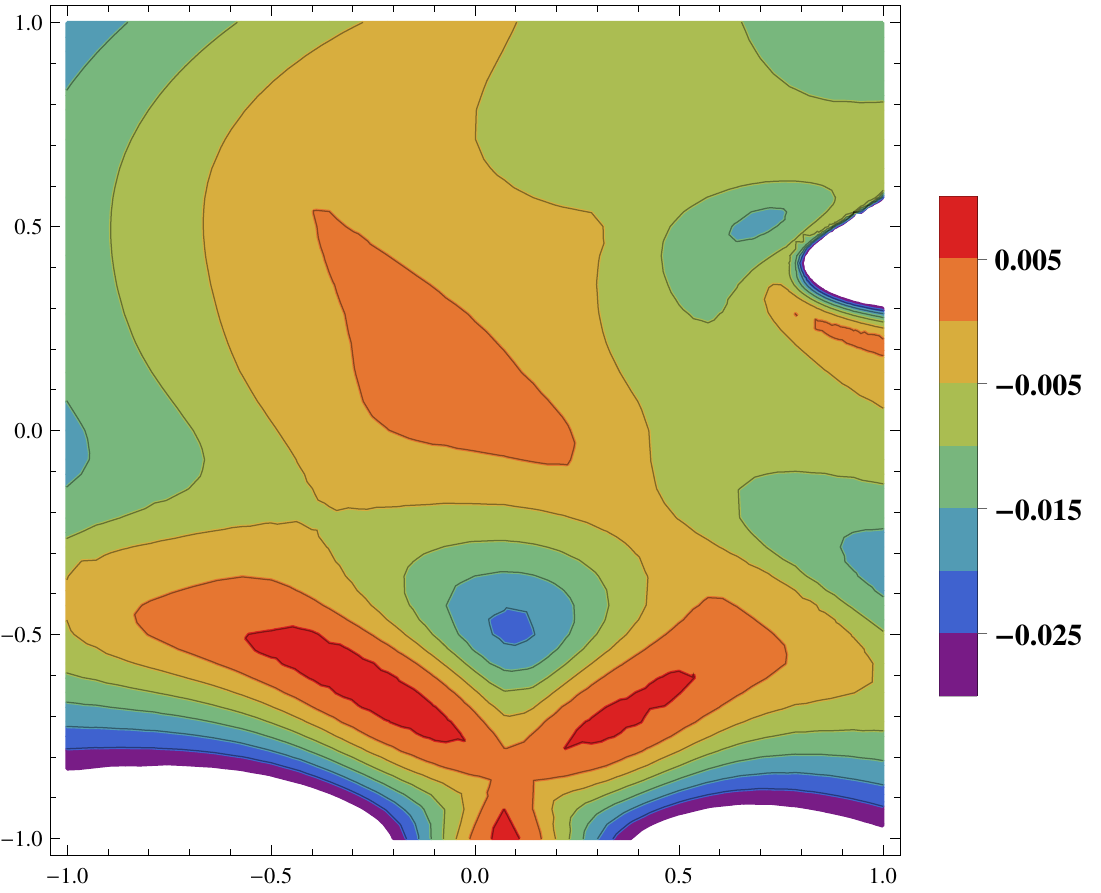}
\includegraphics[scale=.38,keepaspectratio=true]{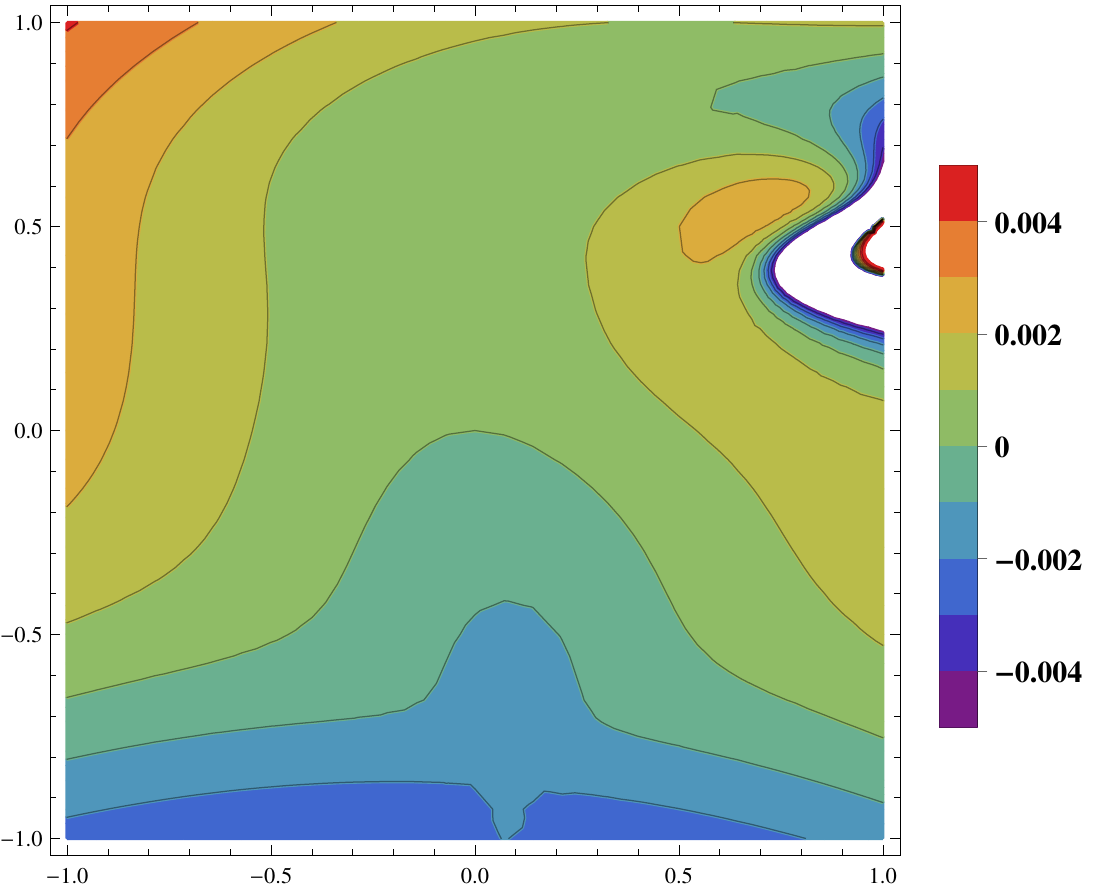}
\caption{The stability (non-normalised mass) (left) and dS (right) landscape of Sol.$~2$. The  solution is located at the origin. The pictures are a 2D slice of the parameters $x = -e(9B_1 -2 B_2 -B_2)/2$ and $y = -e(7A_1 - 2A_2 - A_3)/6$ that are part of a linear combination of  $A_I,B_I$  in eq.~(\ref{eq:SBP}). \label{fig:land}} %
\end{figure}

Finally, it is interesting to consider the interplay between stability and dS solutions. For non-geometric stable dS solutions, the intersection of stability and dS overlap is thin sheets because of small differences in the shape of these landscapes \cite{Danielsson:2012by}, thus requiring fine-tuning. This is not the case for the present non-perturbative solutions. The stability and potential landscapes, plotted in Figure \ref{fig:land}, have noticeably different shapes. This implies that there are sizeable intersection regions\footnote{This fact does not seem to hinge on the duality invariant Ansatz \eqref{W1}; we expect it to hold for more general moduli-dependence.}.
 
This non-trivial overlap will be important when taking flux quantisation into account.
By scaling $N$ large, the parameters $A_I$ and $B_I$ become approximately integers. One can then make $a_i, b_i$ integers by an appropriate truncation.
This will slightly modify the solution, (inversely) related to order to which we  rescale $N$. However, because of the large intersection areas of stability and dS, only a very coarse truncation would significantly modify the solution and possibly spoil stability and/or dS. In our case, where the orientifold tadpole gives a bound on how much rescaling can take place, the truncation would have to be indeed quite coarse. On account of the large stable dS regions, one can achieve quantisation without losing stability nor positive potential energy. We have explicitly checked this for solution 5, which can be rescaled and truncated to the flux parameters
\begin{equation}
\{a_0,a_1,a_2,a_3,b_0,b_1,b_2,b_3\} = \{-1,4,1,-12,4,0,-1,0\}\,,
\end{equation}
which gives a tadpole $N_{O3} = 60$ and rank $K=67$. This has a stable dS solution at 
\bea
\!\!\{ S,T,U \} &\!\approx \!&\{\! .00616 \!+\! i e^{1.32}, - .000456 \!+\! i e^{2.63}, -.117 \!+\! i e^{.0728} \}\nonumber \\
\eea
which is a perturbation of our solution 5.
 \\

\noindent {\bf Discussion}.
In summary, we considered a novel one-step mechanism to stabilise all geometric moduli of type IIB toroidal compactifications in a dS vacuum, using the non-trivial moduli dependence of the tree-level superpotential and the non-perturbative contributions. The latter is motivated by duality invariance of string theory, and can also be seen as a small-field expansion. Our approach  improves the three-step KKLT mechanism  by including  the complex structure in the non-perturbative piece allowing us to stabilise all moduli at once in a dS vacua, avoiding also the introduction of explicitly \cancel{SUSY} terms, such as anti-D-branes. We have presented a number of explicit stable dS solutions, amongst one with quantised fluxes.

We view our results as very compelling arguments to extend the dS landscape in type IIB flux compactifications. They represent a first step towards a new direction allowing for a more complete landscape of stable dS vacua. \\

\smallskip 
\noindent
{\em Acknowledgements.}
The authors would like to thank Ulf Danielsson, Giuseppe Dibitetto, Marc Andre Heller, Thomas Van Riet and Bert Vercnocke for useful discussions. 
JB acknowledges support by the Swedish Research Council (VR) and the G\"oran Gustafsson Foundation.

\begin{table}[htbp!]\tiny
\centering
\input{SBP_table.tex}
\caption{These are the values of the \cancel{SUSY} parameters defined by (\ref{eq:SBP}) that gives the solutions displayed in Table \ref{tab:sols} (rounded to six digits).
\label{tab:SBP}}
\end{table}

\bibliography{refs}

\end{document}

%% file: new_sol_table_Bert.tex
\begin{tabular}{|c||c|c|c|c|c|}
\hline 
 & Sol.~1 & Sol.~2 & Sol.~3 & Sol.~4 & Sol.~5\\
\hline 
$V_0$ & $0.00113$ & $2.23\times 10^{-12}$ & $0.0000251$ & $0.0000234$ & $8.61\times 10^{-12}$\\ 
 \hline 
$\tilde{\gamma}-1$ & $0.0256$ & $5.11 \times 10^{-11}$ & $0.00248$ & $0.0160$ & $6.05 \times 10^{-10}$\\ 
 \hline 
$\frac{|b_i|}{|a_i|}$ & $0.298$ & $0.599$ & $1.32$ & $0.208$ & $0.997$\\ 
 \hline 
$\frac{|\tilde{a}_i|}{|a_i|}$ & $0.611$ & $0.274$ & $0.528$ & $0.621$ & $0.000227$\\ 
 \hline 
Masses & $ \begin{array}{c} 39.0 \\19.7 \\12.4 \\9.74 \\0.00236 \\0.0000747 \\\end{array} $ & $ \begin{array}{c} 2.11\times 10^{10} \\8.71\times 10^9 \\7.00\times 10^9 \\3.41\times 10^9 \\1.26\times 10^9 \\6.01\times 10^8 \\\end{array} $ & $ \begin{array}{c} 1140. \\387. \\106. \\18.4 \\6.16 \\0.0000612 \\\end{array} $ & $ \begin{array}{c} 76.2 \\36.0 \\19.6 \\11.4 \\0.774 \\0.000252 \\\end{array} $ & $ \begin{array}{c} 2.20\times 10^9 \\9.80\times 10^8 \\2.45\times 10^8 \\490000. \\101000. \\100000. \\\end{array} $\\ 
 \hline 
\end{tabular} 

%% file: SBP_table.tex
\begin{tabular}{|c||c|c|c|c|c|}
\hline 
 & Sol.~1 & Sol.~2 & Sol.~3 & Sol.~4 & Sol.~5\\
\hline 
$A_1$ & $-0.147286$ & $-0.0859982$ & $0.0590861$ & $-0.0115235$ & $0.000516097$\\ 
 \hline 
$A_2$ & $0.449418$ & $-1.58993$ & $-0.483429$ & $0.165447$ & $1.15243$\\ 
 \hline 
$A_3$ & $-0.907814$ & $0.4631$ & $-0.131249$ & $-0.144582$ & $-0.000587804$\\ 
 \hline 
$B_1$ & $0.377918$ & $-0.236806$ & $0.0870739$ & $0.0793589$ & $0.00319387$\\ 
 \hline 
$B_2$ & $1.6678$ & $-1.12127$ & $0.826607$ & $0.259372$ & $-0.196848$\\ 
 \hline 
$B_3$ & $0.173821$ & $-0.047207$ & $-0.0614712$ & $0.0902761$ & $-0.00969035$\\ 
 \hline 
\end{tabular} 

%% file: BRZ-PRL-v2.bbl
%merlin.mbs apsrev4-1.bst 2010-07-25 4.21a (PWD, AO, DPC) hacked
%Control: key (0)
%Control: author (8) initials jnrlst
%Control: editor formatted (1) identically to author
%Control: production of article title (-1) disabled
%Control: page (0) single
%Control: year (1) truncated
%Control: production of eprint (0) enabled
\begin{thebibliography}{34}%
\makeatletter
\providecommand \@ifxundefined [1]{%
 \@ifx{#1\undefined}
}%
\providecommand \@ifnum [1]{%
 \ifnum #1\expandafter \@firstoftwo
 \else \expandafter \@secondoftwo
 \fi
}%
\providecommand \@ifx [1]{%
 \ifx #1\expandafter \@firstoftwo
 \else \expandafter \@secondoftwo
 \fi
}%
\providecommand \natexlab [1]{#1}%
\providecommand \enquote  [1]{``#1''}%
\providecommand \bibnamefont  [1]{#1}%
\providecommand \bibfnamefont [1]{#1}%
\providecommand \citenamefont [1]{#1}%
\providecommand \href@noop [0]{\@secondoftwo}%
\providecommand \href [0]{\begingroup \@sanitize@url \@href}%
\providecommand \@href[1]{\@@startlink{#1}\@@href}%
\providecommand \@@href[1]{\endgroup#1\@@endlink}%
\providecommand \@sanitize@url [0]{\catcode `\\12\catcode `\$12\catcode
  `\&12\catcode `\#12\catcode `\^12\catcode `\_12\catcode `\%12\relax}%
\providecommand \@@startlink[1]{}%
\providecommand \@@endlink[0]{}%
\providecommand \url  [0]{\begingroup\@sanitize@url \@url }%
\providecommand \@url [1]{\endgroup\@href {#1}{\urlprefix }}%
\providecommand \urlprefix  [0]{URL }%
\providecommand \Eprint [0]{\href }%
\providecommand \doibase [0]{http://dx.doi.org/}%
\providecommand \selectlanguage [0]{\@gobble}%
\providecommand \bibinfo  [0]{\@secondoftwo}%
\providecommand \bibfield  [0]{\@secondoftwo}%
\providecommand \translation [1]{[#1]}%
\providecommand \BibitemOpen [0]{}%
\providecommand \bibitemStop [0]{}%
\providecommand \bibitemNoStop [0]{.\EOS\space}%
\providecommand \EOS [0]{\spacefactor3000\relax}%
\providecommand \BibitemShut  [1]{\csname bibitem#1\endcsname}%
\let\auto@bib@innerbib\@empty
%</preamble>
\bibitem [{\citenamefont {Kachru}\ \emph {et~al.}(2003)\citenamefont {Kachru},
  \citenamefont {Kallosh}, \citenamefont {Linde},\ and\ \citenamefont
  {Trivedi}}]{KKLT}%
  \BibitemOpen
  \bibfield  {author} {\bibinfo {author} {\bibfnamefont {S.}~\bibnamefont
  {Kachru}}, \bibinfo {author} {\bibfnamefont {R.}~\bibnamefont {Kallosh}},
  \bibinfo {author} {\bibfnamefont {A.~D.}\ \bibnamefont {Linde}}, \ and\
  \bibinfo {author} {\bibfnamefont {S.~P.}\ \bibnamefont {Trivedi}},\ }\href
  {\doibase 10.1103/PhysRevD.68.046005} {\bibfield  {journal} {\bibinfo
  {journal} {Phys.Rev.}\ }\textbf {\bibinfo {volume} {D68}},\ \bibinfo {pages}
  {046005} (\bibinfo {year} {2003})},\ \Eprint
  {http://arxiv.org/abs/hep-th/0301240} {arXiv:hep-th/0301240 [hep-th]}
  \BibitemShut {NoStop}%
%%CITATION = HEP-TH/0301240;%%
\bibitem [{\citenamefont {Giddings}\ \emph {et~al.}(2002)\citenamefont
  {Giddings}, \citenamefont {Kachru},\ and\ \citenamefont {Polchinski}}]{GKP}%
  \BibitemOpen
  \bibfield  {author} {\bibinfo {author} {\bibfnamefont {S.~B.}\ \bibnamefont
  {Giddings}}, \bibinfo {author} {\bibfnamefont {S.}~\bibnamefont {Kachru}}, \
  and\ \bibinfo {author} {\bibfnamefont {J.}~\bibnamefont {Polchinski}},\
  }\href {\doibase 10.1103/PhysRevD.66.106006} {\bibfield  {journal} {\bibinfo
  {journal} {Phys.Rev.}\ }\textbf {\bibinfo {volume} {D66}},\ \bibinfo {pages}
  {106006} (\bibinfo {year} {2002})},\ \Eprint
  {http://arxiv.org/abs/hep-th/0105097} {arXiv:hep-th/0105097 [hep-th]}
  \BibitemShut {NoStop}%
%%CITATION = HEP-TH/0105097;%%
\bibitem [{\citenamefont {DeWolfe}\ \emph {et~al.}(2005)\citenamefont
  {DeWolfe}, \citenamefont {Giryavets}, \citenamefont {Kachru},\ and\
  \citenamefont {Taylor}}]{DeWolfe}%
  \BibitemOpen
  \bibfield  {author} {\bibinfo {author} {\bibfnamefont {O.}~\bibnamefont
  {DeWolfe}}, \bibinfo {author} {\bibfnamefont {A.}~\bibnamefont {Giryavets}},
  \bibinfo {author} {\bibfnamefont {S.}~\bibnamefont {Kachru}}, \ and\ \bibinfo
  {author} {\bibfnamefont {W.}~\bibnamefont {Taylor}},\ }\href {\doibase
  10.1088/1126-6708/2005/07/066} {\bibfield  {journal} {\bibinfo  {journal}
  {JHEP}\ }\textbf {\bibinfo {volume} {0507}},\ \bibinfo {pages} {066}
  (\bibinfo {year} {2005})},\ \Eprint {http://arxiv.org/abs/hep-th/0505160}
  {arXiv:hep-th/0505160 [hep-th]} \BibitemShut {NoStop}%
%%CITATION = HEP-TH/0505160;%%
\bibitem [{\citenamefont {Hertzberg}\ \emph {et~al.}(2007)\citenamefont
  {Hertzberg}, \citenamefont {Kachru}, \citenamefont {Taylor},\ and\
  \citenamefont {Tegmark}}]{Hertzberg}%
  \BibitemOpen
  \bibfield  {author} {\bibinfo {author} {\bibfnamefont {M.~P.}\ \bibnamefont
  {Hertzberg}}, \bibinfo {author} {\bibfnamefont {S.}~\bibnamefont {Kachru}},
  \bibinfo {author} {\bibfnamefont {W.}~\bibnamefont {Taylor}}, \ and\ \bibinfo
  {author} {\bibfnamefont {M.}~\bibnamefont {Tegmark}},\ }\href {\doibase
  10.1088/1126-6708/2007/12/095} {\bibfield  {journal} {\bibinfo  {journal}
  {JHEP}\ }\textbf {\bibinfo {volume} {0712}},\ \bibinfo {pages} {095}
  (\bibinfo {year} {2007})},\ \Eprint {http://arxiv.org/abs/0711.2512}
  {arXiv:0711.2512 [hep-th]} \BibitemShut {NoStop}%
%%CITATION = ARXIV:0711.2512;%%
\bibitem [{\citenamefont {Caviezel}\ \emph {et~al.}(2009)\citenamefont
  {Caviezel}, \citenamefont {Koerber}, \citenamefont {Kors}, \citenamefont
  {Lust}, \citenamefont {Wrase} \emph {et~al.}}]{Munich}%
  \BibitemOpen
  \bibfield  {author} {\bibinfo {author} {\bibfnamefont {C.}~\bibnamefont
  {Caviezel}}, \bibinfo {author} {\bibfnamefont {P.}~\bibnamefont {Koerber}},
  \bibinfo {author} {\bibfnamefont {S.}~\bibnamefont {Kors}}, \bibinfo {author}
  {\bibfnamefont {D.}~\bibnamefont {Lust}}, \bibinfo {author} {\bibfnamefont
  {T.}~\bibnamefont {Wrase}},  \emph {et~al.},\ }\href {\doibase
  10.1088/1126-6708/2009/04/010} {\bibfield  {journal} {\bibinfo  {journal}
  {JHEP}\ }\textbf {\bibinfo {volume} {0904}},\ \bibinfo {pages} {010}
  (\bibinfo {year} {2009})},\ \Eprint {http://arxiv.org/abs/0812.3551}
  {arXiv:0812.3551 [hep-th]} \BibitemShut {NoStop}%
%%CITATION = ARXIV:0812.3551;%%
\bibitem [{\citenamefont {Flauger}\ \emph {et~al.}(2009)\citenamefont
  {Flauger}, \citenamefont {Paban}, \citenamefont {Robbins},\ and\
  \citenamefont {Wrase}}]{Flauger:2008ad}%
  \BibitemOpen
  \bibfield  {author} {\bibinfo {author} {\bibfnamefont {R.}~\bibnamefont
  {Flauger}}, \bibinfo {author} {\bibfnamefont {S.}~\bibnamefont {Paban}},
  \bibinfo {author} {\bibfnamefont {D.}~\bibnamefont {Robbins}}, \ and\
  \bibinfo {author} {\bibfnamefont {T.}~\bibnamefont {Wrase}},\ }\href
  {\doibase 10.1103/PhysRevD.79.086011} {\bibfield  {journal} {\bibinfo
  {journal} {Phys.Rev.}\ }\textbf {\bibinfo {volume} {D79}},\ \bibinfo {pages}
  {086011} (\bibinfo {year} {2009})},\ \Eprint {http://arxiv.org/abs/0812.3886}
  {arXiv:0812.3886 [hep-th]} \BibitemShut {NoStop}%
%%CITATION = ARXIV:0812.3886;%%
\bibitem [{\citenamefont {Witten}(1996)}]{Witten:1996bn}%
  \BibitemOpen
  \bibfield  {author} {\bibinfo {author} {\bibfnamefont {E.}~\bibnamefont
  {Witten}},\ }\href {\doibase 10.1016/0550-3213(96)00283-0} {\bibfield
  {journal} {\bibinfo  {journal} {Nucl.Phys.}\ }\textbf {\bibinfo {volume}
  {B474}},\ \bibinfo {pages} {343} (\bibinfo {year} {1996})},\ \Eprint
  {http://arxiv.org/abs/hep-th/9604030} {arXiv:hep-th/9604030 [hep-th]}
  \BibitemShut {NoStop}%
%%CITATION = HEP-TH/9604030;%%
\bibitem [{\citenamefont {de~Alwis}(2005{\natexlab{a}})}]{deAlwis:2005tf}%
  \BibitemOpen
  \bibfield  {author} {\bibinfo {author} {\bibfnamefont {S.}~\bibnamefont
  {de~Alwis}},\ }\href {\doibase 10.1016/j.physletb.2005.08.096} {\bibfield
  {journal} {\bibinfo  {journal} {Phys.Lett.}\ }\textbf {\bibinfo {volume}
  {B626}},\ \bibinfo {pages} {223} (\bibinfo {year} {2005}{\natexlab{a}})},\
  \Eprint {http://arxiv.org/abs/hep-th/0506266} {arXiv:hep-th/0506266 [hep-th]}
  \BibitemShut {NoStop}%
%%CITATION = HEP-TH/0506266;%%
\bibitem [{\citenamefont {de~Alwis}(2005{\natexlab{b}})}]{Shanta1}%
  \BibitemOpen
  \bibfield  {author} {\bibinfo {author} {\bibfnamefont {S.}~\bibnamefont
  {de~Alwis}},\ }\href {\doibase 10.1016/j.physletb.2005.09.027} {\bibfield
  {journal} {\bibinfo  {journal} {Phys.Lett.}\ }\textbf {\bibinfo {volume}
  {B628}},\ \bibinfo {pages} {183} (\bibinfo {year} {2005}{\natexlab{b}})},\
  \Eprint {http://arxiv.org/abs/hep-th/0506267} {arXiv:hep-th/0506267 [hep-th]}
  \BibitemShut {NoStop}%
%%CITATION = HEP-TH/0506267;%%
\bibitem [{\citenamefont {Choi}\ \emph {et~al.}(2004)\citenamefont {Choi},
  \citenamefont {Falkowski}, \citenamefont {Nilles}, \citenamefont
  {Olechowski},\ and\ \citenamefont {Pokorski}}]{Choi:2004sx}%
  \BibitemOpen
  \bibfield  {author} {\bibinfo {author} {\bibfnamefont {K.}~\bibnamefont
  {Choi}}, \bibinfo {author} {\bibfnamefont {A.}~\bibnamefont {Falkowski}},
  \bibinfo {author} {\bibfnamefont {H.~P.}\ \bibnamefont {Nilles}}, \bibinfo
  {author} {\bibfnamefont {M.}~\bibnamefont {Olechowski}}, \ and\ \bibinfo
  {author} {\bibfnamefont {S.}~\bibnamefont {Pokorski}},\ }\href {\doibase
  10.1088/1126-6708/2004/11/076} {\bibfield  {journal} {\bibinfo  {journal}
  {JHEP}\ }\textbf {\bibinfo {volume} {0411}},\ \bibinfo {pages} {076}
  (\bibinfo {year} {2004})},\ \Eprint {http://arxiv.org/abs/hep-th/0411066}
  {arXiv:hep-th/0411066 [hep-th]} \BibitemShut {NoStop}%
%%CITATION = HEP-TH/0411066;%%
\bibitem [{\citenamefont {Gallego}\ and\ \citenamefont
  {Serone}(2009)}]{Gallego}%
  \BibitemOpen
  \bibfield  {author} {\bibinfo {author} {\bibfnamefont {D.}~\bibnamefont
  {Gallego}}\ and\ \bibinfo {author} {\bibfnamefont {M.}~\bibnamefont
  {Serone}},\ }\href {\doibase 10.1088/1126-6708/2009/01/056} {\bibfield
  {journal} {\bibinfo  {journal} {JHEP}\ }\textbf {\bibinfo {volume} {0901}},\
  \bibinfo {pages} {056} (\bibinfo {year} {2009})},\ \Eprint
  {http://arxiv.org/abs/0812.0369} {arXiv:0812.0369 [hep-th]} \BibitemShut
  {NoStop}%
%%CITATION = ARXIV:0812.0369;%%
\bibitem [{\citenamefont {Burgess}\ \emph {et~al.}(2003)\citenamefont
  {Burgess}, \citenamefont {Kallosh},\ and\ \citenamefont
  {Quevedo}}]{Burgess:2003ic}%
  \BibitemOpen
  \bibfield  {author} {\bibinfo {author} {\bibfnamefont {C.}~\bibnamefont
  {Burgess}}, \bibinfo {author} {\bibfnamefont {R.}~\bibnamefont {Kallosh}}, \
  and\ \bibinfo {author} {\bibfnamefont {F.}~\bibnamefont {Quevedo}},\
  }\href@noop {} {\bibfield  {journal} {\bibinfo  {journal} {JHEP}\ }\textbf
  {\bibinfo {volume} {0310}},\ \bibinfo {pages} {056} (\bibinfo {year}
  {2003})},\ \Eprint {http://arxiv.org/abs/hep-th/0309187}
  {arXiv:hep-th/0309187 [hep-th]} \BibitemShut {NoStop}%
%%CITATION = HEP-TH/0309187;%%
\bibitem [{\citenamefont {Achucarro}\ \emph {et~al.}(2006)\citenamefont
  {Achucarro}, \citenamefont {de~Carlos}, \citenamefont {Casas},\ and\
  \citenamefont {Doplicher}}]{Achucarro:2006zf}%
  \BibitemOpen
  \bibfield  {author} {\bibinfo {author} {\bibfnamefont {A.}~\bibnamefont
  {Achucarro}}, \bibinfo {author} {\bibfnamefont {B.}~\bibnamefont
  {de~Carlos}}, \bibinfo {author} {\bibfnamefont {J.}~\bibnamefont {Casas}}, \
  and\ \bibinfo {author} {\bibfnamefont {L.}~\bibnamefont {Doplicher}},\ }\href
  {\doibase 10.1088/1126-6708/2006/06/014} {\bibfield  {journal} {\bibinfo
  {journal} {JHEP}\ }\textbf {\bibinfo {volume} {0606}},\ \bibinfo {pages}
  {014} (\bibinfo {year} {2006})},\ \Eprint
  {http://arxiv.org/abs/hep-th/0601190} {arXiv:hep-th/0601190 [hep-th]}
  \BibitemShut {NoStop}%
%%CITATION = HEP-TH/0601190;%%
\bibitem [{\citenamefont {Covi}\ \emph {et~al.}(2009)\citenamefont {Covi},
  \citenamefont {Gomez-Reino}, \citenamefont {Gross}, \citenamefont {Palma},\
  and\ \citenamefont {Scrucca}}]{Marta1}%
  \BibitemOpen
  \bibfield  {author} {\bibinfo {author} {\bibfnamefont {L.}~\bibnamefont
  {Covi}}, \bibinfo {author} {\bibfnamefont {M.}~\bibnamefont {Gomez-Reino}},
  \bibinfo {author} {\bibfnamefont {C.}~\bibnamefont {Gross}}, \bibinfo
  {author} {\bibfnamefont {G.~A.}\ \bibnamefont {Palma}}, \ and\ \bibinfo
  {author} {\bibfnamefont {C.~A.}\ \bibnamefont {Scrucca}},\ }\href {\doibase
  10.1088/1126-6708/2009/03/146} {\bibfield  {journal} {\bibinfo  {journal}
  {JHEP}\ }\textbf {\bibinfo {volume} {0903}},\ \bibinfo {pages} {146}
  (\bibinfo {year} {2009})},\ \Eprint {http://arxiv.org/abs/0812.3864}
  {arXiv:0812.3864 [hep-th]} \BibitemShut {NoStop}%
%%CITATION = ARXIV:0812.3864;%%
\bibitem [{\citenamefont {Cicoli}\ \emph
  {et~al.}(2012{\natexlab{a}})\citenamefont {Cicoli}, \citenamefont {Maharana},
  \citenamefont {Quevedo},\ and\ \citenamefont {Burgess}}]{Fer}%
  \BibitemOpen
  \bibfield  {author} {\bibinfo {author} {\bibfnamefont {M.}~\bibnamefont
  {Cicoli}}, \bibinfo {author} {\bibfnamefont {A.}~\bibnamefont {Maharana}},
  \bibinfo {author} {\bibfnamefont {F.}~\bibnamefont {Quevedo}}, \ and\
  \bibinfo {author} {\bibfnamefont {C.}~\bibnamefont {Burgess}},\ }\href
  {\doibase 10.1007/JHEP06(2012)011} {\bibfield  {journal} {\bibinfo  {journal}
  {JHEP}\ }\textbf {\bibinfo {volume} {1206}},\ \bibinfo {pages} {011}
  (\bibinfo {year} {2012}{\natexlab{a}})},\ \Eprint
  {http://arxiv.org/abs/1203.1750} {arXiv:1203.1750 [hep-th]} \BibitemShut
  {NoStop}%
%%CITATION = ARXIV:1203.1750;%%
\bibitem [{\citenamefont {Louis}\ \emph {et~al.}(2012)\citenamefont {Louis},
  \citenamefont {Rummel}, \citenamefont {Valandro},\ and\ \citenamefont
  {Westphal}}]{Louis}%
  \BibitemOpen
  \bibfield  {author} {\bibinfo {author} {\bibfnamefont {J.}~\bibnamefont
  {Louis}}, \bibinfo {author} {\bibfnamefont {M.}~\bibnamefont {Rummel}},
  \bibinfo {author} {\bibfnamefont {R.}~\bibnamefont {Valandro}}, \ and\
  \bibinfo {author} {\bibfnamefont {A.}~\bibnamefont {Westphal}},\ }\href
  {\doibase 10.1007/JHEP10(2012)163} {\bibfield  {journal} {\bibinfo  {journal}
  {JHEP}\ }\textbf {\bibinfo {volume} {1210}},\ \bibinfo {pages} {163}
  (\bibinfo {year} {2012})},\ \Eprint {http://arxiv.org/abs/1208.3208}
  {arXiv:1208.3208 [hep-th]} \BibitemShut {NoStop}%
%%CITATION = ARXIV:1208.3208;%%
\bibitem [{\citenamefont {Parameswaran}\ \emph {et~al.}(2011)\citenamefont
  {Parameswaran}, \citenamefont {Ramos-Sanchez},\ and\ \citenamefont
  {Zavala}}]{Parameswaran:2010ec}%
  \BibitemOpen
  \bibfield  {author} {\bibinfo {author} {\bibfnamefont {S.~L.}\ \bibnamefont
  {Parameswaran}}, \bibinfo {author} {\bibfnamefont {S.}~\bibnamefont
  {Ramos-Sanchez}}, \ and\ \bibinfo {author} {\bibfnamefont {I.}~\bibnamefont
  {Zavala}},\ }\href {\doibase 10.1007/JHEP01(2011)071} {\bibfield  {journal}
  {\bibinfo  {journal} {JHEP}\ }\textbf {\bibinfo {volume} {1101}},\ \bibinfo
  {pages} {071} (\bibinfo {year} {2011})},\ \Eprint
  {http://arxiv.org/abs/1009.3931} {arXiv:1009.3931 [hep-th]} \BibitemShut
  {NoStop}%
%%CITATION = ARXIV:1009.3931;%%
\bibitem [{\citenamefont {Balasubramanian}\ \emph {et~al.}(2005)\citenamefont
  {Balasubramanian}, \citenamefont {Berglund}, \citenamefont {Conlon},\ and\
  \citenamefont {Quevedo}}]{Balasubramanian}%
  \BibitemOpen
  \bibfield  {author} {\bibinfo {author} {\bibfnamefont {V.}~\bibnamefont
  {Balasubramanian}}, \bibinfo {author} {\bibfnamefont {P.}~\bibnamefont
  {Berglund}}, \bibinfo {author} {\bibfnamefont {J.~P.}\ \bibnamefont
  {Conlon}}, \ and\ \bibinfo {author} {\bibfnamefont {F.}~\bibnamefont
  {Quevedo}},\ }\href {\doibase 10.1088/1126-6708/2005/03/007} {\bibfield
  {journal} {\bibinfo  {journal} {JHEP}\ }\textbf {\bibinfo {volume} {0503}},\
  \bibinfo {pages} {007} (\bibinfo {year} {2005})},\ \Eprint
  {http://arxiv.org/abs/hep-th/0502058} {arXiv:hep-th/0502058 [hep-th]}
  \BibitemShut {NoStop}%
%%CITATION = HEP-TH/0502058;%%
\bibitem [{\citenamefont {Cicoli}\ \emph
  {et~al.}(2012{\natexlab{b}})\citenamefont {Cicoli}, \citenamefont
  {Krippendorf}, \citenamefont {Mayrhofer}, \citenamefont {Quevedo},\ and\
  \citenamefont {Valandro}}]{LV1}%
  \BibitemOpen
  \bibfield  {author} {\bibinfo {author} {\bibfnamefont {M.}~\bibnamefont
  {Cicoli}}, \bibinfo {author} {\bibfnamefont {S.}~\bibnamefont {Krippendorf}},
  \bibinfo {author} {\bibfnamefont {C.}~\bibnamefont {Mayrhofer}}, \bibinfo
  {author} {\bibfnamefont {F.}~\bibnamefont {Quevedo}}, \ and\ \bibinfo
  {author} {\bibfnamefont {R.}~\bibnamefont {Valandro}},\ }\href {\doibase
  10.1007/JHEP09(2012)019} {\bibfield  {journal} {\bibinfo  {journal} {JHEP}\
  }\textbf {\bibinfo {volume} {1209}},\ \bibinfo {pages} {019} (\bibinfo {year}
  {2012}{\natexlab{b}})},\ \Eprint {http://arxiv.org/abs/1206.5237}
  {arXiv:1206.5237 [hep-th]} \BibitemShut {NoStop}%
%%CITATION = ARXIV:1206.5237;%%
\bibitem [{\citenamefont {Cicoli}\ \emph
  {et~al.}(2013{\natexlab{a}})\citenamefont {Cicoli}, \citenamefont
  {Krippendorf}, \citenamefont {Mayrhofer}, \citenamefont {Quevedo},\ and\
  \citenamefont {Valandro}}]{LV2}%
  \BibitemOpen
  \bibfield  {author} {\bibinfo {author} {\bibfnamefont {M.}~\bibnamefont
  {Cicoli}}, \bibinfo {author} {\bibfnamefont {S.}~\bibnamefont {Krippendorf}},
  \bibinfo {author} {\bibfnamefont {C.}~\bibnamefont {Mayrhofer}}, \bibinfo
  {author} {\bibfnamefont {F.}~\bibnamefont {Quevedo}}, \ and\ \bibinfo
  {author} {\bibfnamefont {R.}~\bibnamefont {Valandro}},\ }\href {\doibase
  10.1007/JHEP07(2013)150} {\bibfield  {journal} {\bibinfo  {journal} {JHEP}\
  }\textbf {\bibinfo {volume} {1307}},\ \bibinfo {pages} {150} (\bibinfo {year}
  {2013}{\natexlab{a}})},\ \Eprint {http://arxiv.org/abs/1304.0022}
  {arXiv:1304.0022 [hep-th]} \BibitemShut {NoStop}%
%%CITATION = ARXIV:1304.0022;%%
\bibitem [{\citenamefont {Cicoli}\ \emph
  {et~al.}(2013{\natexlab{b}})\citenamefont {Cicoli}, \citenamefont {Klevers},
  \citenamefont {Krippendorf}, \citenamefont {Mayrhofer}, \citenamefont
  {Quevedo} \emph {et~al.}}]{LV3}%
  \BibitemOpen
  \bibfield  {author} {\bibinfo {author} {\bibfnamefont {M.}~\bibnamefont
  {Cicoli}}, \bibinfo {author} {\bibfnamefont {D.}~\bibnamefont {Klevers}},
  \bibinfo {author} {\bibfnamefont {S.}~\bibnamefont {Krippendorf}}, \bibinfo
  {author} {\bibfnamefont {C.}~\bibnamefont {Mayrhofer}}, \bibinfo {author}
  {\bibfnamefont {F.}~\bibnamefont {Quevedo}},  \emph {et~al.},\ }\href@noop {}
  {\  (\bibinfo {year} {2013}{\natexlab{b}})},\ \Eprint
  {http://arxiv.org/abs/1312.0014} {arXiv:1312.0014 [hep-th]} \BibitemShut
  {NoStop}%
%%CITATION = ARXIV:1312.0014;%%
\bibitem [{\citenamefont {Danielsson}\ and\ \citenamefont
  {Dibitetto}(2013{\natexlab{a}})}]{Danielsson:2013XX}%
  \BibitemOpen
  \bibfield  {author} {\bibinfo {author} {\bibfnamefont {U.}~\bibnamefont
  {Danielsson}}\ and\ \bibinfo {author} {\bibfnamefont {G.}~\bibnamefont
  {Dibitetto}},\ }\href@noop {} {\  (\bibinfo {year} {2013}{\natexlab{a}})},\
  \Eprint {http://arxiv.org/abs/1312.5331} {arXiv:1312.5331 [hep-th]}
  \BibitemShut {NoStop}%
%%CITATION = ARXIV:1312.5331;%%
\bibitem [{\citenamefont {Shelton}\ \emph {et~al.}(2005)\citenamefont
  {Shelton}, \citenamefont {Taylor},\ and\ \citenamefont {Wecht}}]{Wecht}%
  \BibitemOpen
  \bibfield  {author} {\bibinfo {author} {\bibfnamefont {J.}~\bibnamefont
  {Shelton}}, \bibinfo {author} {\bibfnamefont {W.}~\bibnamefont {Taylor}}, \
  and\ \bibinfo {author} {\bibfnamefont {B.}~\bibnamefont {Wecht}},\ }\href
  {\doibase 10.1088/1126-6708/2005/10/085} {\bibfield  {journal} {\bibinfo
  {journal} {JHEP}\ }\textbf {\bibinfo {volume} {0510}},\ \bibinfo {pages}
  {085} (\bibinfo {year} {2005})},\ \Eprint
  {http://arxiv.org/abs/hep-th/0508133} {arXiv:hep-th/0508133 [hep-th]}
  \BibitemShut {NoStop}%
%%CITATION = HEP-TH/0508133;%%
\bibitem [{\citenamefont {de~Carlos}\ \emph {et~al.}(2010)\citenamefont
  {de~Carlos}, \citenamefont {Guarino},\ and\ \citenamefont
  {Moreno}}]{Guarino}%
  \BibitemOpen
  \bibfield  {author} {\bibinfo {author} {\bibfnamefont {B.}~\bibnamefont
  {de~Carlos}}, \bibinfo {author} {\bibfnamefont {A.}~\bibnamefont {Guarino}},
  \ and\ \bibinfo {author} {\bibfnamefont {J.~M.}\ \bibnamefont {Moreno}},\
  }\href {\doibase 10.1007/JHEP02(2010)076} {\bibfield  {journal} {\bibinfo
  {journal} {JHEP}\ }\textbf {\bibinfo {volume} {1002}},\ \bibinfo {pages}
  {076} (\bibinfo {year} {2010})},\ \Eprint {http://arxiv.org/abs/0911.2876}
  {arXiv:0911.2876 [hep-th]} \BibitemShut {NoStop}%
%%CITATION = ARXIV:0911.2876;%%
\bibitem [{\citenamefont {Uranga}(2009)}]{Uranga}%
  \BibitemOpen
  \bibfield  {author} {\bibinfo {author} {\bibfnamefont {A.~M.}\ \bibnamefont
  {Uranga}},\ }\href {\doibase 10.1088/1126-6708/2009/01/048} {\bibfield
  {journal} {\bibinfo  {journal} {JHEP}\ }\textbf {\bibinfo {volume} {0901}},\
  \bibinfo {pages} {048} (\bibinfo {year} {2009})},\ \Eprint
  {http://arxiv.org/abs/0808.2918} {arXiv:0808.2918 [hep-th]} \BibitemShut
  {NoStop}%
%%CITATION = ARXIV:0808.2918;%%
\bibitem [{\citenamefont {Dibitetto}\ \emph {et~al.}(2011)\citenamefont
  {Dibitetto}, \citenamefont {Guarino},\ and\ \citenamefont
  {Roest}}]{Dibitetto:2011gm}%
  \BibitemOpen
  \bibfield  {author} {\bibinfo {author} {\bibfnamefont {G.}~\bibnamefont
  {Dibitetto}}, \bibinfo {author} {\bibfnamefont {A.}~\bibnamefont {Guarino}},
  \ and\ \bibinfo {author} {\bibfnamefont {D.}~\bibnamefont {Roest}},\ }\href
  {\doibase 10.1007/JHEP03(2011)137} {\bibfield  {journal} {\bibinfo  {journal}
  {JHEP}\ }\textbf {\bibinfo {volume} {1103}},\ \bibinfo {pages} {137}
  (\bibinfo {year} {2011})},\ \Eprint {http://arxiv.org/abs/1102.0239}
  {arXiv:1102.0239 [hep-th]} \BibitemShut {NoStop}%
%%CITATION = ARXIV:1102.0239;%%
\bibitem [{\citenamefont {Dall'Agata}\ and\ \citenamefont
  {Inverso}(2012)}]{DallAgata}%
  \BibitemOpen
  \bibfield  {author} {\bibinfo {author} {\bibfnamefont {G.}~\bibnamefont
  {Dall'Agata}}\ and\ \bibinfo {author} {\bibfnamefont {G.}~\bibnamefont
  {Inverso}},\ }\href {\doibase 10.1016/j.nuclphysb.2012.01.023} {\bibfield
  {journal} {\bibinfo  {journal} {Nucl.Phys.}\ }\textbf {\bibinfo {volume}
  {B859}},\ \bibinfo {pages} {70} (\bibinfo {year} {2012})},\ \Eprint
  {http://arxiv.org/abs/1112.3345} {arXiv:1112.3345 [hep-th]} \BibitemShut
  {NoStop}%
%%CITATION = ARXIV:1112.3345;%%
\bibitem [{\citenamefont {Borghese}\ \emph {et~al.}(2012)\citenamefont
  {Borghese}, \citenamefont {Guarino},\ and\ \citenamefont {Roest}}]{Borghese}%
  \BibitemOpen
  \bibfield  {author} {\bibinfo {author} {\bibfnamefont {A.}~\bibnamefont
  {Borghese}}, \bibinfo {author} {\bibfnamefont {A.}~\bibnamefont {Guarino}}, \
  and\ \bibinfo {author} {\bibfnamefont {D.}~\bibnamefont {Roest}},\ }\href
  {\doibase 10.1007/JHEP12(2012)108} {\bibfield  {journal} {\bibinfo  {journal}
  {JHEP}\ }\textbf {\bibinfo {volume} {1212}},\ \bibinfo {pages} {108}
  (\bibinfo {year} {2012})},\ \Eprint {http://arxiv.org/abs/1209.3003}
  {arXiv:1209.3003 [hep-th]} \BibitemShut {NoStop}%
%%CITATION = ARXIV:1209.3003;%%
\bibitem [{\citenamefont {Danielsson}\ and\ \citenamefont
  {Dibitetto}(2013{\natexlab{b}})}]{Danielsson:2012by}%
  \BibitemOpen
  \bibfield  {author} {\bibinfo {author} {\bibfnamefont {U.}~\bibnamefont
  {Danielsson}}\ and\ \bibinfo {author} {\bibfnamefont {G.}~\bibnamefont
  {Dibitetto}},\ }\href {\doibase 10.1007/JHEP03(2013)018} {\bibfield
  {journal} {\bibinfo  {journal} {JHEP}\ }\textbf {\bibinfo {volume} {1303}},\
  \bibinfo {pages} {018} (\bibinfo {year} {2013}{\natexlab{b}})},\ \Eprint
  {http://arxiv.org/abs/1212.4984} {arXiv:1212.4984 [hep-th]} \BibitemShut
  {NoStop}%
%%CITATION = ARXIV:1212.4984;%%
\bibitem [{\citenamefont {Bl{\aa}b{\"a}ck}\ \emph
  {et~al.}(2013{\natexlab{a}})\citenamefont {Bl{\aa}b{\"a}ck}, \citenamefont
  {Danielsson},\ and\ \citenamefont {Dibitetto}}]{Blaback:2013ht}%
  \BibitemOpen
  \bibfield  {author} {\bibinfo {author} {\bibfnamefont {J.}~\bibnamefont
  {Bl{\aa}b{\"a}ck}}, \bibinfo {author} {\bibfnamefont {U.}~\bibnamefont
  {Danielsson}}, \ and\ \bibinfo {author} {\bibfnamefont {G.}~\bibnamefont
  {Dibitetto}},\ }\href {\doibase 10.1007/JHEP08(2013)054} {\bibfield
  {journal} {\bibinfo  {journal} {JHEP}\ }\textbf {\bibinfo {volume} {1308}},\
  \bibinfo {pages} {054} (\bibinfo {year} {2013}{\natexlab{a}})},\ \Eprint
  {http://arxiv.org/abs/1301.7073} {arXiv:1301.7073 [hep-th]} \BibitemShut
  {NoStop}%
%%CITATION = ARXIV:1301.7073;%%
\bibitem [{\citenamefont {Damian}\ \emph {et~al.}(2013)\citenamefont {Damian},
  \citenamefont {Diaz-Barron}, \citenamefont {Loaiza-Brito},\ and\
  \citenamefont {Sabido}}]{Damian:2013dq}%
  \BibitemOpen
  \bibfield  {author} {\bibinfo {author} {\bibfnamefont {C.}~\bibnamefont
  {Damian}}, \bibinfo {author} {\bibfnamefont {L.~R.}\ \bibnamefont
  {Diaz-Barron}}, \bibinfo {author} {\bibfnamefont {O.}~\bibnamefont
  {Loaiza-Brito}}, \ and\ \bibinfo {author} {\bibfnamefont {M.}~\bibnamefont
  {Sabido}},\ }\href {\doibase 10.1007/JHEP06(2013)109} {\bibfield  {journal}
  {\bibinfo  {journal} {JHEP}\ }\textbf {\bibinfo {volume} {1306}},\ \bibinfo
  {pages} {109} (\bibinfo {year} {2013})},\ \Eprint
  {http://arxiv.org/abs/1302.0529} {arXiv:1302.0529 [hep-th]} \BibitemShut
  {NoStop}%
%%CITATION = ARXIV:1302.0529;%%
\bibitem [{\citenamefont {Damian}\ and\ \citenamefont
  {Loaiza-Brito}(2013)}]{Damian:2013dwa}%
  \BibitemOpen
  \bibfield  {author} {\bibinfo {author} {\bibfnamefont {C.}~\bibnamefont
  {Damian}}\ and\ \bibinfo {author} {\bibfnamefont {O.}~\bibnamefont
  {Loaiza-Brito}},\ }\href {\doibase 10.1103/PhysRevD.88.046008} {\bibfield
  {journal} {\bibinfo  {journal} {Phys.Rev.}\ }\textbf {\bibinfo {volume}
  {D88}},\ \bibinfo {pages} {046008} (\bibinfo {year} {2013})},\ \Eprint
  {http://arxiv.org/abs/1304.0792} {arXiv:1304.0792 [hep-th]} \BibitemShut
  {NoStop}%
%%CITATION = ARXIV:1304.0792;%%
\bibitem [{\citenamefont {Bl{\aa}b{\"a}ck}\ \emph
  {et~al.}(2013{\natexlab{b}})\citenamefont {Bl{\aa}b{\"a}ck}, \citenamefont
  {Danielsson},\ and\ \citenamefont {Dibitetto}}]{Blaback:2013fca}%
  \BibitemOpen
  \bibfield  {author} {\bibinfo {author} {\bibfnamefont {J.}~\bibnamefont
  {Bl{\aa}b{\"a}ck}}, \bibinfo {author} {\bibfnamefont {U.}~\bibnamefont
  {Danielsson}}, \ and\ \bibinfo {author} {\bibfnamefont {G.}~\bibnamefont
  {Dibitetto}},\ }\href@noop {} {\  (\bibinfo {year} {2013}{\natexlab{b}})},\
  \Eprint {http://arxiv.org/abs/1310.8300} {arXiv:1310.8300 [hep-th]}
  \BibitemShut {NoStop}%
%%CITATION = ARXIV:1310.8300;%%
\bibitem [{\citenamefont {Gukov}\ \emph {et~al.}(2000)\citenamefont {Gukov},
  \citenamefont {Vafa},\ and\ \citenamefont {Witten}}]{GVW}%
  \BibitemOpen
  \bibfield  {author} {\bibinfo {author} {\bibfnamefont {S.}~\bibnamefont
  {Gukov}}, \bibinfo {author} {\bibfnamefont {C.}~\bibnamefont {Vafa}}, \ and\
  \bibinfo {author} {\bibfnamefont {E.}~\bibnamefont {Witten}},\ }\href
  {\doibase 10.1016/S0550-3213(00)00373-4} {\bibfield  {journal} {\bibinfo
  {journal} {Nucl.Phys.}\ }\textbf {\bibinfo {volume} {B584}},\ \bibinfo
  {pages} {69} (\bibinfo {year} {2000})},\ \Eprint
  {http://arxiv.org/abs/hep-th/9906070} {arXiv:hep-th/9906070 [hep-th]}
  \BibitemShut {NoStop}%
%%CITATION = HEP-TH/9906070;%%
\end{thebibliography}%
